\documentclass[12pt]{iopart}

\usepackage{graphicx}
\usepackage{caption}
\usepackage{subcaption}
\begin{document}

\title[Energetic particle transport in optimized stellarators
]{Energetic particle transport in optimized stellarators}

\author{A Bader$^1$, D T Anderson$^1$, M Drevlak$^2$, B J Faber$^1$, C C Hegna$^1$, S Henneberg$^2$, M Landreman$^3$, J C Schmitt$^4$, Y Suzuki$^5$, A. Ware$^6$}

\address{1: University of Wisconsin-Madison, Madison, WI, USA\\ 2: Max-Planck-Institut für Plasmaphysik, Greifswald, Germany\\ 3: University of Maryland, College Park, MD, USA\\ 4: Auburn University, Auburn, AL, USA\\ 5: National Institute for Fusion Science, Toki, Japan \\ 6: University of Montana, Missoula, MT, USA Montana}
\ead{abader@engr.wisc.edu}
\vspace{10pt}

\begin{abstract}
  Nine stellarator configurations, three quasiaxisymmetric, three quasihelically symmetric and three non-quasisymmetric are scaled to ARIES-CS size and analyzed for energetic particle content.  The best performing configurations with regard to energetic particle confinement also perform the best on the neoclassical $\Gamma_c$ metric, which attempts to align contours of the second adiabatic invariant with flux surfaces. Quasisymmetric configurations that simultaneously perform well on $\Gamma_c$ and quasisymmetry have the best overall confinement, with collisional losses under 3\%, approaching the performance of ITER with ferritic inserts. 
\end{abstract}

%
%
%
%
%

\section{Introduction}

Confining energetic particles, especially alpha particles born in nuclear fusion reactions, is of key importance for magnetic confinement fusion reactors.  In configurations where axisymmetry is not present, either tokamaks with non-axisymmetric perturbations, or stellarators, some particle orbits will have a non-zero bounce-averaged radial drift causing them to leave the confined region, sometimes very quickly.  These promptly lost particles can cause significant damage to plasma facing surfaces and reduce the lifetime of the plasma wall \cite{mau2008divertor}. It will never be possible to confine all alpha particles in a fusion device, but reducing the losses, especially prompt losses, is crucial for the longevity of the device.  This paper will show collisional energetic particle transport results from various stellarator configurations at the reactor scale with the goal of identifying the properties of configurations with good confinement.

Several metrics have been developed for neo-classical confinement in 3D systems \cite{grieger1992physics}. Some configurations possess a symmetry in the magnetic field strength, $|B|$, and therefore are isomorphic to axisymmetric systems.  These configurations are called quasi-symmetric, because they possess a symmetry in $|B|$ similar to an axisymmetric system \cite{reiman1999physics}.  This paper includes configurations of two quasi-symmetric types, quasi-axisymmetric (QA) where $|B|$ contours connect toroidally, and quasi-helical (QH) where $|B|$ contours connect helically \cite{Nuhrenberg_PLA_1988}. In all the configurations exact quasisymmetry is not present, but rather an approximate symmetry exists.  The deviation from a strict symmetry is referred to in this paper as quasisymmetric deviation.  Mathematically this is obtained by transforming coordinates into the Boozer coordinate system, and then calculating the energy in the non-symmetric modes.

It has been shown experimentally that quasi-helical configurations improve neoclassical confinement \cite{canik2007experimental}. Additionally, many numerical explorations of quasisymmetric configurations of all types exist as well. Specifically, recent results indicate that low quasi-symmetric deviation can help alpha confinement in both quasiaxisymmetric \cite{henneberg2019improving} and quasi-helically symmetric configurations \cite{bader2019stellarator}.

There are alternative methods for improving neoclassical confinement in the absence of quasisymmetry exist. A broader class of configurations are omnigenous, in which the second adiabatic invariant, $J_\parallel = \oint v_\parallel dl$, is constant on a flux surface. A consequence of this optimization is that all the maxima and minima of a field line on a flux surface are the same. Configurations that approximate omnigeneity are called quasi-omnigenous (QO). If, in addition, flux surfaces close poloidally, the configuration is quasi-isodynamic. A consequence of this optimization is that drifts are purely poloidal, and bootstrap and Pfirsch-Schl\"uter currents vanish. This optimization was used to produce W7-X \cite{nuhrenberg2010development}. Confinement of energetic particles in quasi-isodynamic configurations is expected to improve at high pressure when the alignment between $J_\parallel$ and flux surfaces improves \cite{lotz1992collisionless}.

An even less restrictive optimization for improved confinement is described as $\sigma$-optimization \cite{mynick1983improved}. In LHD it is possible to achieve an equilibria where $\sigma = 1$, where for a given flux surface the minima of each field line are equivalent, but the maxima do not. This is achieved in LHD by shifting the axis inward, creating the "inward shifted" configuration \cite{murakami2002neoclassical}. In this optimization, collisionless drift orbits are not fully confined, but the coefficient of neoclassical transport drops significantly.

Various metrics have been used to quantify the degree of neoclassical transport. One metric, $\epsilon_\mathrm{eff}$, is the coefficient of the neoclassical diffusion in the low-collisionality ($\sim 1/\nu$) regime has been regularly used for stellarator optimization \cite{nemov1999evaluation}. However, recent results indicate that there is little correlation between $\epsilon_\mathrm{eff}$ and good energetic particle confinement \cite{bader2019stellarator}.  However, a different metric $\Gamma_c$ \cite{Nemov_PoP_2005, Nemov_PoP_2008}, that seeks explicitly to align contours of $J_\parallel$ with flux surfaces similar to the omnigeneity constraint, has been shown to correlate better with energetic particle confinement. This metric has been used to optimize quasi-helically symmetric configurations with good energetic particle confinement \cite{bader2019stellarator, bader2020advancing}.

In previous publications stellarators were compared only between variations of similar classes (QA, QH, QO). Comparisons between stellarators of different classes, have often been hampered by different choices of magnetic field strength, size, and profiles of density and temperature. Some calculations include collisions where others do not. Comparisons between published results is therefore very difficult. This paper attempts to rectify the situation by providing consistent scalings across a broad class of configurations and then comparing the energetic particle confinements both collisionlessly and with collisions.

The layout of this paper is as follows. In Section \ref{sec:configs}, we will briefly describe the configurations used in this paper. Section \ref{sec:scaling} will explain how the reactor scale configurations were constructed.  Section \ref{sec:coll} will show results from both collisional and collisionless calculations of alpha particles.  Section \ref{sec:metrics} compares the alpha particle losses for the metrics of interest in each configuration. Section \ref{sec:disc} will discuss the results and describe the limitations of the current work. Section \ref{sec:conc} will conclude the paper and provide areas for future research.

\section{Configurations}
\label{sec:configs}

This paper considers three quasi-helically symmetric configurations, three quasi-axisymmetric configurations, a W7-X like configuration, and two LHD-like configurations. An ITER configuration is included for comparison.  A table of all the configurations and their relevant properties has been included in table \ref{tab:configs}.

\begin{table}
\caption{\label{jlab1}A list of configurations along with relevant properties}
\footnotesize
\begin{tabular}{@{}lllll}
\br
Name&Type&Periods&Aspect ratio&$\beta$\\
\mr
Wistell-A & QH & 4 & 6.7 & Vacuum\\
Wistell-B & QH & 5 & 6.6 & Vacuum\\
Ku5 & QH & 4 & 10.0 & 10.0\%\\
ARIES-CS & QA & 3 & 4.5 & 4.0\%\\
NCSX & QA & 3 & 4.4 & 4.3\%\\
Simsopt & QA & 2 & 6.0 & Vacuum\\
W7-X & QI & 5 & 10.5 & 4.4\%\\
LHD st & Torsotron & 10 & 6.5 & Vacuum\\
LHD in. & Torsotron & 10 & 6.2 &Vacuum\\
ITER&Tokamak&N/A&2.5&2.2\%\\
\br
\label{tab:configs}
\end{tabular}\\
\end{table}
\normalsize

The quasi-helically symmetric configurations are: the "Wistell-A" configuration which has been described in a previous publication \cite{bader2020advancing}; the "Wistell-B" configuration, a five-field period vacuum configuration optimized with the \texttt{ROSE} code \cite{Drevlak_NF_2018} explicitly for quasisymmetry and $\Gamma_c$; "Ku4" a four field period configuration from \cite{ku2010new} that was optimized for quasisymmetry at high normalized pressure, $\beta$. The three quasi-axisymmetric configurations are comprised of the NCSX (specifically "li383") \cite{koniges2003magnetic, mynick2002exploration} and ARIES-CS (specifically: "n3are") \cite{ku2008physics, mynick2006improving} configurations. Also included amongst quasiaxisymmetric configurations is a more recent vacuum configuration. called Simsopt, which was optimized solely for quasisymmetry at the $s=0.5$ surface using the \texttt{SIMSOPT} optimizer \cite{simsoptcode}.  Here and throughout the paper, $s$ represents the normalized toroidal flux, $\psi/\psi_{\mathrm{edge}}$.  The W7-X like configuration is a high-mirror configuration designed for improved energetic particle confinement \cite{grieger1991physics}, with coefficients given by Table IV in \cite{nuhrenberg1996global}. The two LHD \cite{murakami2002neoclassical} configurations are vacuum configurations, with one in the standard (outward) configuration and the other in an inward shifted configuration which is known to have improved confinement properties \cite{mynick1983improved}.  Finally the ITER configuration is a near-axisymmetric configuration, although this equilibrium includes coil ripple, blanket modules and ferritic inserts \cite{tobita2003reduction}.

\section{Scaling to Reactor size}
\label{sec:scaling}

In order to properly scale configurations to each other it is necessary to adjust to a benchmark size. The ARIES-CS parameters are used for the scaling, representing a fairly compact reactor size. There are two possible ways to scale the configurations.  One option is to scale the configurations to have the same volume (444 m$^3$) the other is to scale the minor radii to the same value (1.7m).  For this paper we only show results using the volume scaling, although the main conclusions do not change when the configurations are scaled to have equivalent minor radii.

\begin{figure}
    \centering
    \begin{subfigure}{0.25\textwidth}
    \centering
    \includegraphics[width=\linewidth]{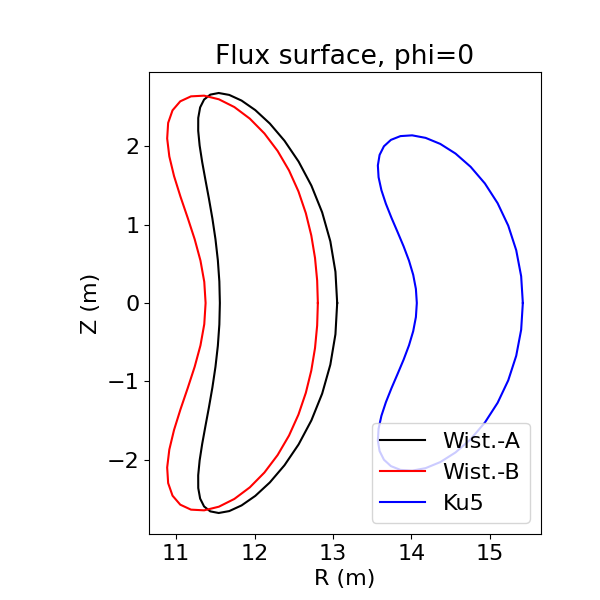}    
    \end{subfigure}%
    \begin{subfigure}{0.32\textwidth}
    \centering
    \includegraphics[width=\linewidth]{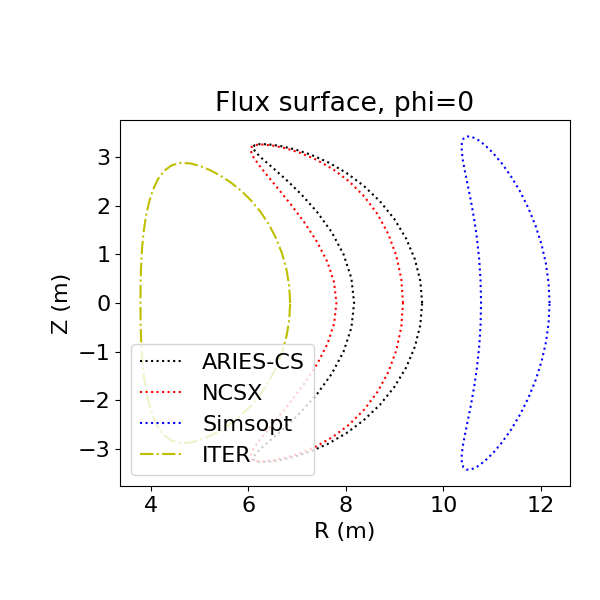}    
    \end{subfigure}%
    \begin{subfigure}{0.43\textwidth}
    \centering
    \includegraphics[width=\linewidth]{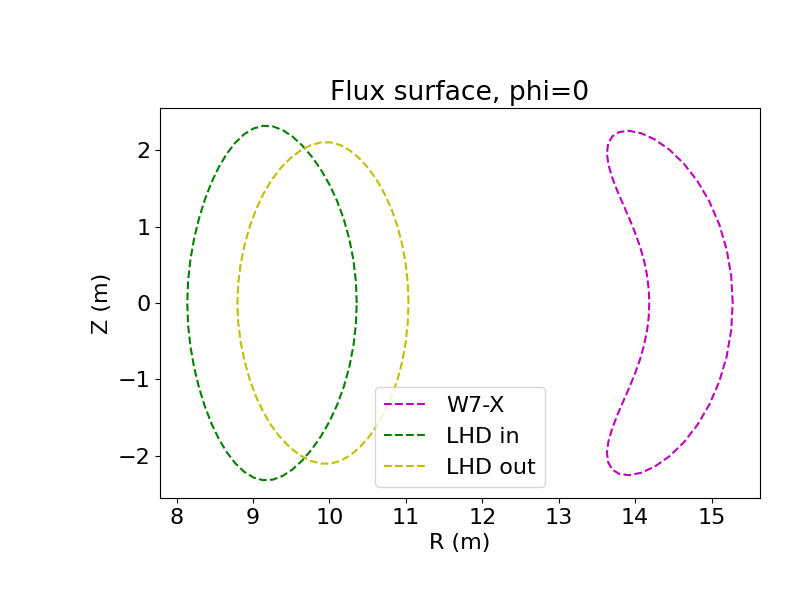}    
    \end{subfigure}%
    \caption{Boundary flux surfaces for the scaled configurations are shown. Left: Quasihelically-symmetric configurations. Center: Quasiaxisymmetric configurations (including ITER) Right: non-quasisymmetric configurations}
    \label{fig:fs}
\end{figure}

All configurations are represented by \texttt{VMEC} equilibria \cite{Hirshman_POF_1983}, and the size scaling is accomplished by adjusting the boundary coefficients such that all configurations have the same volume. The magnetic field strengths are made equivalent by ensuring the volume averaged magnetic field is equivalent across all configurations (5.86 T). For non-vacuum configurations, the rotational transform are adjusted for each configuration by using $I \propto RB$, where $I$ is the total current in the plasma. The normalized pressure, $\beta$ is similarly kept constant through the scaling procedure. The boundary flux surfaces for the configurations at the $\phi$=0 plane, often referred to as the "bean" or "crescent", are plotted in figure \ref{fig:fs}

For all configurations, with three exceptions, the scaling is accomplished by starting from an idealized fixed boundary equilibrium and scaling the coefficients. These fixed boundary equilibria are usually generated through optimization and do not include effects from finite coils. One exception is the ITER equilibrium which includes coil ripple and the effect from ferritic inserts and blanket modules. The scaled ITER equilibrium is a direct replica of the unscaled ITER equilibrium, but with slightly larger volume and higher field. There was no attempt to recalculate the effects of blanket modules and ferritic inserts on the larger size. The other exceptions are the two LHD equilibria which represent two configurations very similar to those generated in the actual LHD device. In these cases, the coils used to generate the equilibria were adjusted and new free-boundary equilibria were generated from the enlarged coils. Results from these free-boundary equilibria are presented in this paper.  A second set of LHD calculations were undertaken with a scaled fixed-boundary equilibrium and the differences were not noticeable, and are not included here.

\begin{figure}
    \centering
    \begin{subfigure}{0.5\textwidth}
    \centering
    \includegraphics[width=\linewidth]{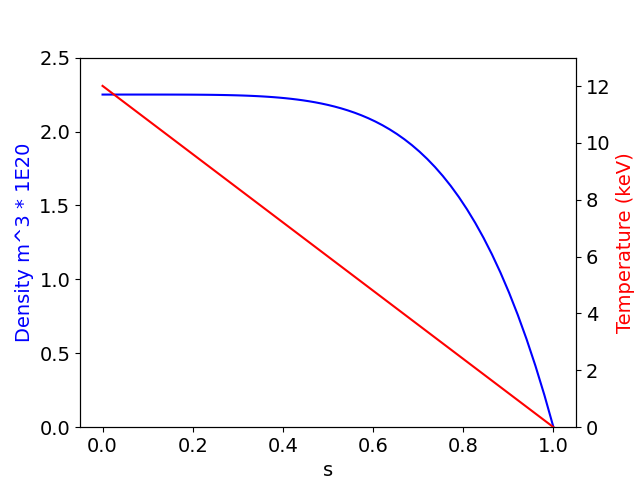}    
    \end{subfigure}%
    \begin{subfigure}{0.5\textwidth}
    \centering
    \includegraphics[width=\linewidth]{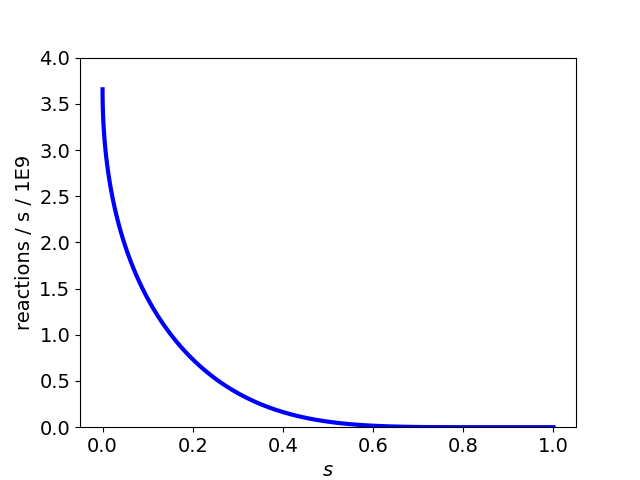}    
    \end{subfigure}%
    \caption{Left: Temperature (red) and density (blue) profiles as a function of normalized toroidal flux $s$. Right: The derived reaction rate given as a function of temperature and density for the ARIES-CS equilibrium.}
    \label{fig:profiles}
\end{figure}

In order to perform the collisional calculations, it is necessary to define the temperature and density profiles. These profiles determine the initial launch points for collisional calculations as well as the slowing down behavior of the alpha particles. The density profile chosen for these simulations is mostly flat with $n = n_0 \left(1-s^5\right)$, the temperature profile is more peaked with $T = T_0\left(1-s\right)$. These profiles are roughly consistent with those chosen for the ARIES-CS studies \cite{ku2008physics}. In the previous equations, $s$ represents the normalized toroidal flux, $\psi/\psi_{edge}$.  The values of the core temperatures, $T_0$ and $n_0$ are approximately equivalent to those of ARIES-CS: $n_{e,0} = 4.8 \times 10^{20}$ and $n_{D,0} = n_{T,0} = 2.25 \times 10^{20}$, with $T_{e,0} = T_{i,0} = 11.5$ keV.  The difference between $n_D + n_T$ and $n_e$ arises from a flat profile of $Z_\mathrm{eff} = 1.13$ as in the ARIES-CS equilibrium. However, collisions with impurity ions are not included in these calculations. Once the temperature profiles are chosen, the reaction profile is determined. The temperature, density and reaction profiles are shown in figure \ref{fig:profiles}. The same reaction profile is used for each equilibrium, and is estimated as

{\setlength{\mathindent}{1cm}
\begin{equation}
    R \frac{dV}{ds} = n_D n_T \langle \sigma v \rangle \frac{dV}{ds};\; \langle \sigma v \rangle = 3.6\times10^{-18} T^{-2/3} \exp{\left(-19.94 * T^{-1/3}\right)} \mathrm{m}^3/\mathrm{sec} 
\end{equation}}

where $n_D$ and $n_T$ are the deuterium and tritium concentrations, $T$ is the temperature in keV and $dV/ds$ is the derivative of the volume with respect to normalized toroidal flux $s$ for the ARIES-CS equilibrium. Even though the reaction profile varies slightly from equilibrium to equilibrium due to variations in $dV/ds$, the same fusion reaction profile is maintained in order to keep consistent particle launch profiles across equilibria.

Since the configurations vary in pressure from vacuum configurations to normalized pressures of 10\%, it is impossible to choose profiles consistent across configurations that also match the pressure profiles from each configuration.  The main goal is to determine what magnetic configuration properties affect alpha particle confinement rather than to do self-consistent studies of each of the configurations. Therefore, the same temperature and density profiles are used in all configurations for alpha particle confinement calculations even though there is no self-consistency with the plasma pressure used in the equilibrium.

\section{Alpha Particle Losses}
\label{sec:coll}

Particles are sourced by first choosing a radial location such that the distribution matches the reaction profile given in figure \ref{fig:profiles}. Next a random location on the surface and the velocity pitch angle is chosen in the same manner as described in \cite{bader2019stellarator}. The guiding centers of the particles are followed using an Adams-Bashford integration scheme and can under go both slowing down and pitch angle scattering. The \texttt{ANTS} code is used for all particle following calculations \cite{Drevlak_NF_2014}. If a particle passes beyond the penultimate flux surface at any point in time it is considered lost. If the particle's energy is the same as the background thermal particles it is considered confined and is no longer followed.

\begin{figure}
    \centering
    \includegraphics[width=\linewidth]{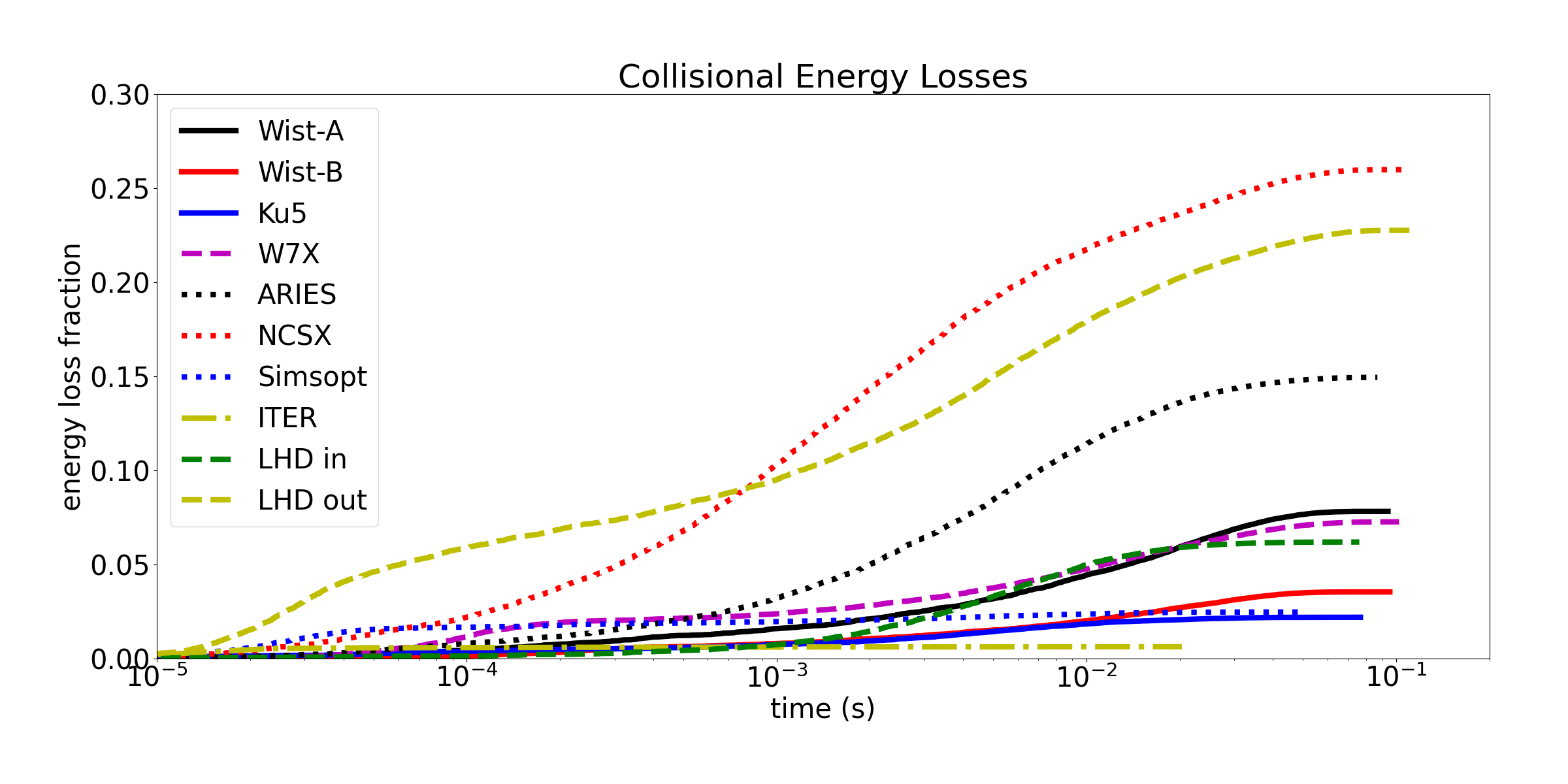}
    \caption{Energy loss from alpha particles as a function of time for all configurations. }
    \label{fig:eloss}
\end{figure}

The results from the collisional calculation are shown in Figure \ref{fig:eloss}. The line style indicates the configuration type, with solid lines indicating QH, dotted lines QA, dashed lines for the LHD-like configurations and both ITER and W7-X use dashed-dotted lines. To help the reader, throughout the paper consistent colors and linestyles (where possible) for each configuration are used. Among the quasisymmetric configuration, the QHs strongly outperform the QAs with the exception of the Simsopt configuration which performs as well as the best QHs.  The three best performing configurations shown (outside of ITER) are the Ku5 configuration (QH), the Wistell-B configuration (QH) and the Simsopt configuration (QA). W7-X performs about equivalently to both the WISTELL-A configuration and the inward shifted LHD configuration. This behavior will be examined in depth later. Note that due to differences in machine size, magnetic field, and particle sourcing, the results shown here may differ from previously published results on energetic particle confinement.

\begin{figure}
    \centering
    \includegraphics[width=\linewidth]{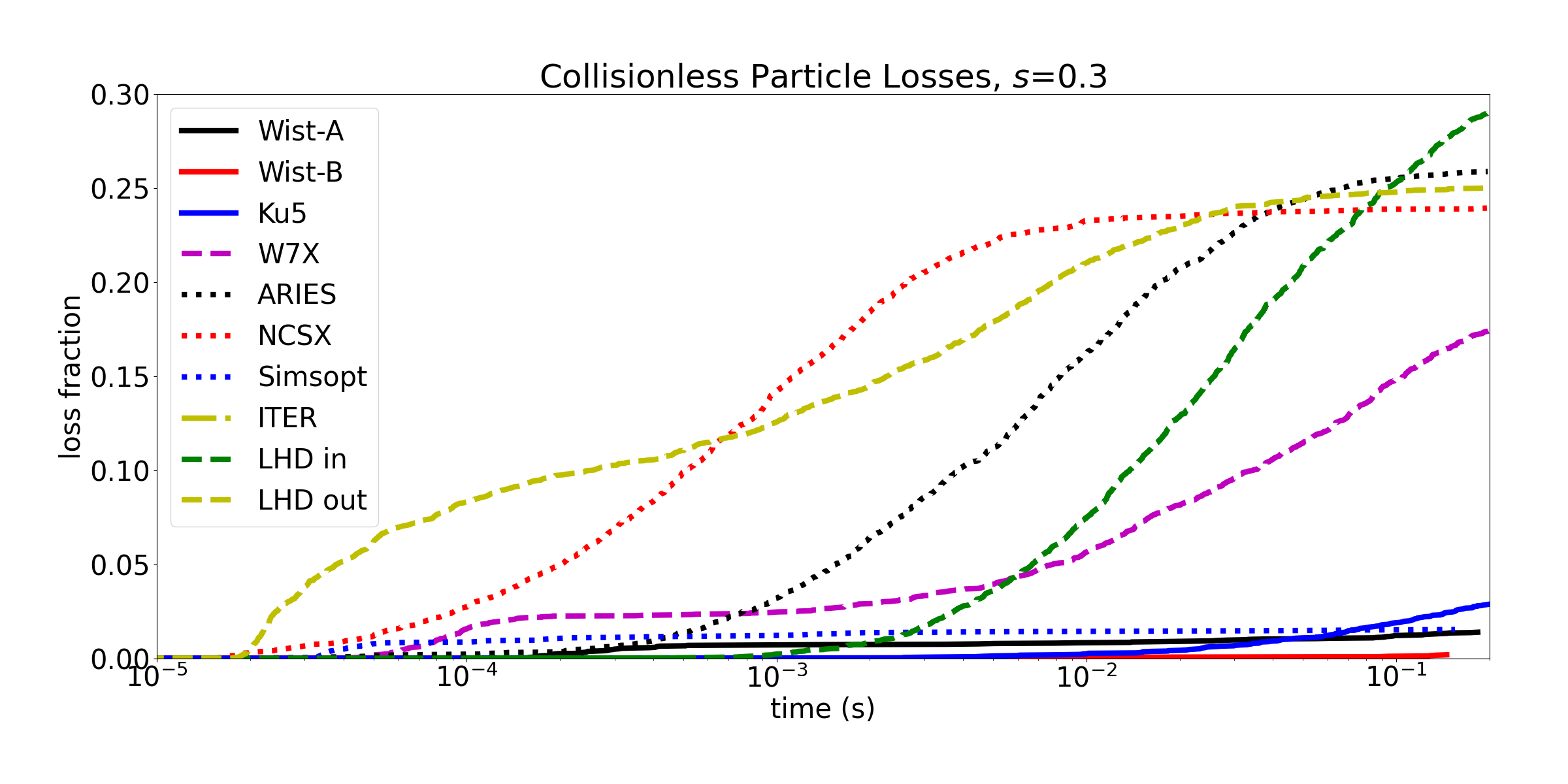}
    \caption{Particle loss as a function of time for all configurations for particles born on the $s$=0.3 surface.}
    \label{fig:ploss}
\end{figure}

In addition to the collisional calculation, calculations without collisions are also presented in Figure \ref{fig:ploss}. For these calculations particles were started on a specific flux surface, in this case $s$=0.3, representing a surface just outside the midradius is chosen. Particles are launched on this surface and followed until they are lost or 200 ms have elapsed, corresponding to several ($\sim3$) slowing down times. Figure \ref{fig:ploss} shows the particle loss versus time rather than the energy loss, but because no collisions are included, all lost particles have the full energy. Collisionless calculations were previously used to distinguish between configurations \cite{bader2019stellarator}, and they are very useful to highlight the specific loss behaviors of the configurations, which will be examined below. Note that the addition of collisions tend to enhance energetic particle losses due to pitch angle scattering onto lost orbits.

Before a more detailed look at the configurations, it will be useful to distinguish between the different types of losses seen. Some particles are born on lost orbits and leave the confined region at almost the full energy values even in the collisional calculation. Particles born on the outer regions of the plasma are likely to be lost in this manner. In fact all the losses from ITER are particles born near the edge that are promptly lost. We will refer to these particles as ``prompt" losses.  

There is a second class of lost particles that undergo many orbits before being lost. Sometimes this is the result from diffusive properties, especially pitch-angle scattering which becomes increasingly important at low energies. However, these slow losses can also occur in collisionless calculations discussed below. As such we will refer to all losses on extended time scales as "stochastic" losses, following the convention in \cite{albert2020accelerated}.

The exact boundary between prompt and stochastic losses is not clear in all configurations, but it is often easy to see the distinction in some of the configurations. The W7-X collisionless losses at, say, $s = 0.3$ are particularly clear. The W7-X configuration loses about 2\% of launched particles born at $s=0.3$ before 0.2 ms. There are almost no additional losses until about 1 ms when additional stochastic losses begin accumulating again. Many of the particles are lost stochastically, however the precise behavior is important. Slow stochastic losses are less problematic because particles will be able to deposit most of their energy. The same distinction between prompt and stochastic losses in W7-X exists with collisionless losses on other flux surfaces (not shown). It also is visible in the collisional losses, however, with collisional losses, diffusive behavior causes there to be some particles losses between 0.2 and 1 ms. A more detailed discussion about prompt versus stochastic losses is in sections \ref{ss:lhd} and \ref{ss:qh}. 

\section{Alpha Particle Loss Metrics}\
\label{sec:metrics}

We consider two metrics for alpha particle losses, quasisymmetry and $\Gamma_c$.  A given configuration is quasisymmetric if the variation of $|B|$ along a field line is the same for all field lines on a flux surface \cite{helander2014theory}. Quasisymmetry can be determined by Fourier decomposing $|B|$ on a flux surface in the straight field line coordinate system known as Boozer coordinates \cite{Boozer_PoF_1982}. If the only modes present are ones where the ratio of the toroidal mode $n$ to the poloidal mode $m$ is constant, the configuration is quasisymmetric. For quasiaxisymmetric equilibria, $n/m = 0$ and a perfectly quasiaxisymmetric equilibrium will only have modes with $n=0$.  For quasihelically-symmetric equilibria, the ratio $n/m$ is usually equal to the number of field periods (modulo a sign). So for a perfectly quasihelically-symmetric equilibrium with four periods, the only modes present are ones where $n/m$ = 4. A third symmetry, quasipoloidal symmetry, where only modes with $m=0$ are present, is not considered in this paper.

Excepting precisely axisymmetric configurations, it is conjectured that perfect quasisymmetry can only be achieved on a single flux surface \cite{garren1991existence}, and a metric is needed to describe the deviation from perfect quasisymmetry. The metric used in this paper is calculated by first Fourier decomposing the two dimensional flux surface in Boozer coordinates, and then summing the magnetic energy in all non-symmetric modes normalized to the $m=0, n=0$ mode, which is representative of the background field strength.  That is, 
\begin{equation}
    Q_{qs}(s) = \frac{1}{B_{0,0}(s)}\left(\sum_{m/n \neq C_{qs}} B_{m,n}^2(s) \right)^{1/2}  
\end{equation}

where $C_{qs}$ represents the target for quasisymmetry, 0 for quasiaxisymmetry and the number of field periods for quasihelical symmetry. Lower values of $Q_{qs}$ indicate better quasisymmetry.

\begin{figure}
    \centering
    \begin{subfigure}{0.5\textwidth}
    \centering
    \includegraphics[width=\linewidth]{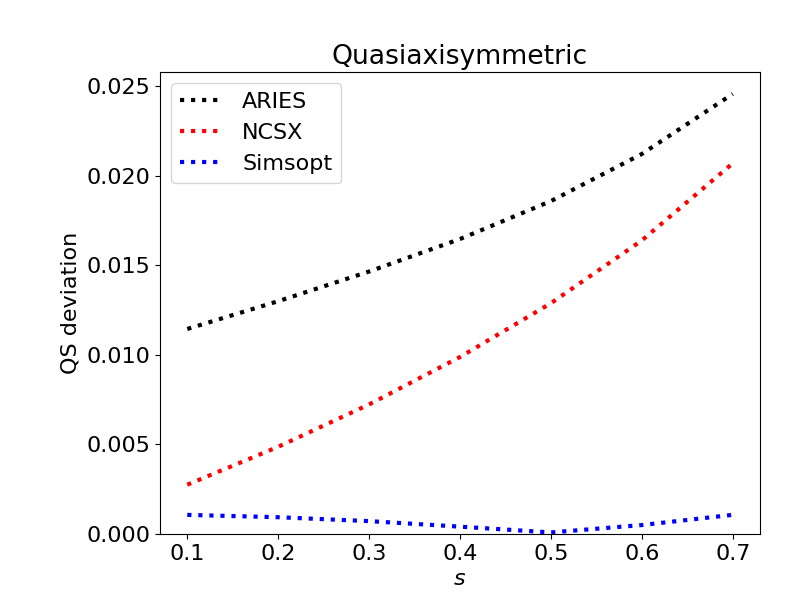}    
    \end{subfigure}%
    \begin{subfigure}{0.5\textwidth}
    \centering
    \includegraphics[width=\linewidth]{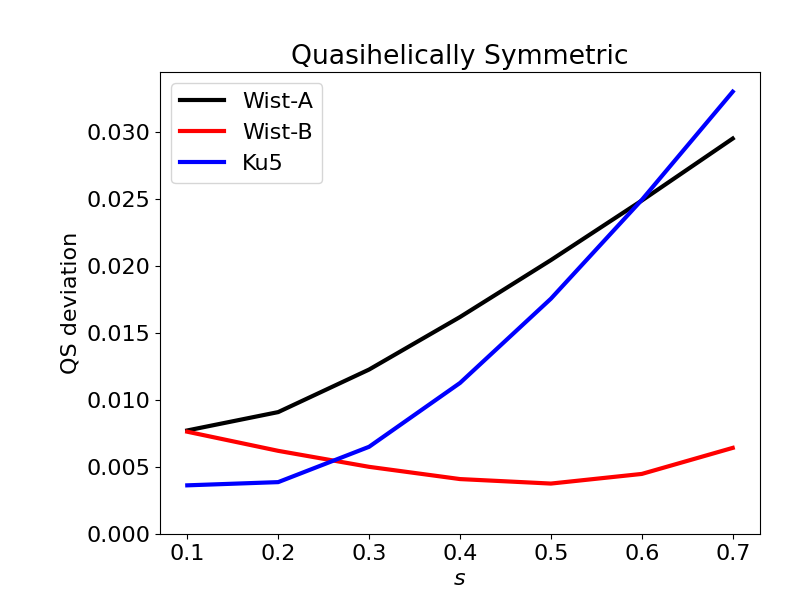}    
    \end{subfigure}%
    \caption{The deviation from quasisymmetry as a function of normalized toroidal flux, $s$ for quasiaxisymmetric configurations (left) and quasihelically symmetric configurations (right)}
    \label{fig:qsqx}
\end{figure}

Figure \ref{fig:qsqx} shows the results of $Q_{qs}$ for quasiaxisymmetric configurations (left) and quasihelically symmetric configurations (right) as a function of flux surface. There is clear separation among the quasiaxisymmetric configurations. At all $s$ values, the ARIES-CS configuration is the least quasisymmetric and the Simsopt configuration is the most quasisymmetric. The story is less clear for the quasihelical configurations. The Ku5 configuration is the most quasisymmetric in the core and the least quasisymmetric in the edge. Overall the Wistell-B configuration has the best average quasisymmetry.

The second metric, $\Gamma_c$ was introduced by Nemov \cite{Nemov_PoP_2008} (eqs 61, 50, and 36) as a measure of the energetic ion confinement properties and is given by,
\begin{equation}
        \Gamma_c = \frac{\pi}{\sqrt{8}} \lim_{L_s \rightarrow \infty} \left( \int_0^{L_s} \frac{ds}{B}\right)^{-1}
        \int_1^{B_\mathrm{max}/B_\mathrm{min}} db' \sum_\mathrm{well_j} \gamma_c^2 \frac{v \tau_{b,j}}{4 B_\mathrm{min}
        b'^{2}};\;\gamma_c = \frac{2}{\pi} \mathrm{arctan}\frac{v_r}{v_\theta}
\end{equation}

Here, $v_r$ and $v_\theta$ are the bounce average radial and poloidal drifts respectively; $v$ is the particle velocity; $\tau_b$ is the bounce time, $B_\mathrm{max}$ and $B_\mathrm{min}$ are the maximum and minimum field strength on a flux surface or suitably long field line; $b'$ represents a normalized field strength, here equivalent to $|B|/B_\mathrm{min}$ and $L_s$ is the length along a field line. The summation is over every well along a field line, where the boundaries of the wells are themselves a function of the integrating variable, $b'$. When $\Gamma_c$ is small, contours of $J_\parallel$ align with flux surfaces, and the bounce average radial drift goes to zero. More information about $\Gamma_c$ and its use for stellarator optimization can be found in \cite{bader2019stellarator}. Unlike quasisymmetry, the $\Gamma_c$ metric can be calculated for all stellarator configurations. All calculations for both $\Gamma_c$ and quasisymmetry were carried out using the \texttt{ROSE} code. Due to an unresolved difficulty with handling single field period equilibria, $\Gamma_c$ for the ITER calculation is unavailable for this paper.

\begin{figure}
    \centering
    \begin{subfigure}{0.5\textwidth}
    \centering
    \includegraphics[width=\linewidth]{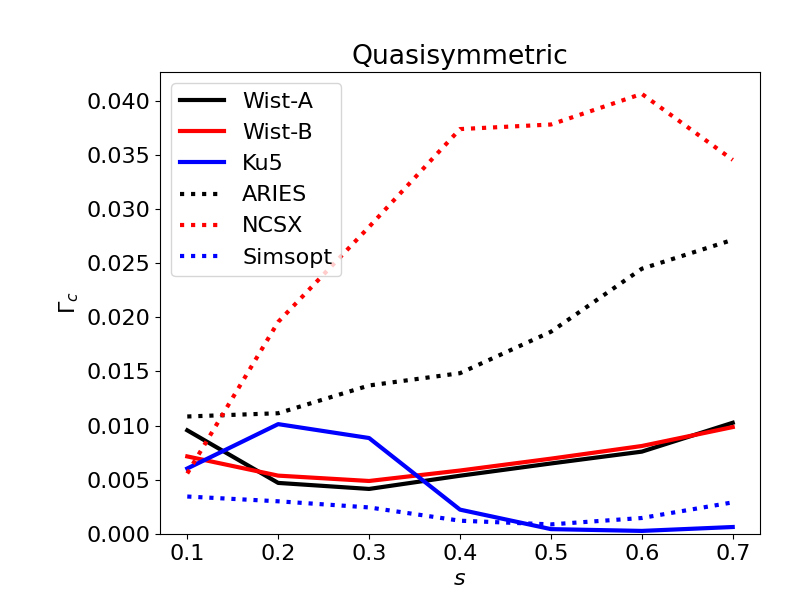}    
    \end{subfigure}%
    \begin{subfigure}{0.5\textwidth}
    \centering
    \includegraphics[width=\linewidth]{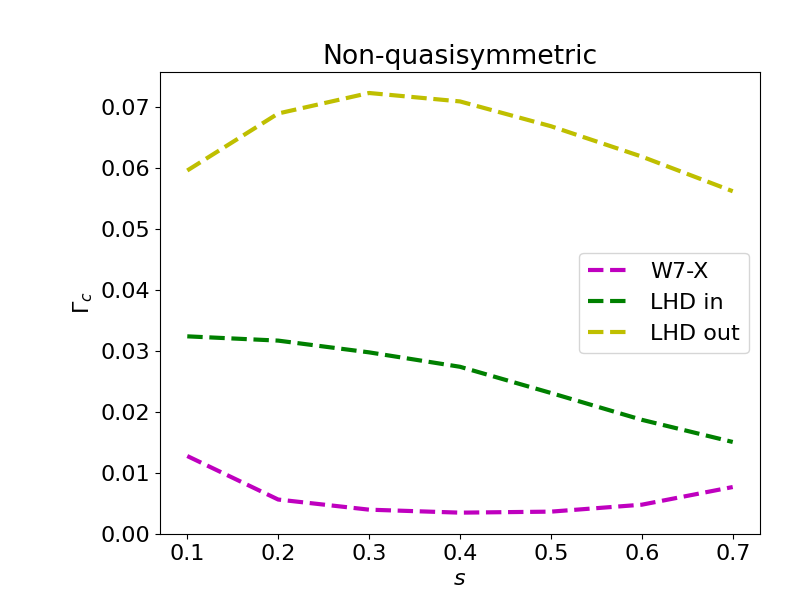}    
    \end{subfigure}%
    \caption{$\Gamma_c$ as a function of normalized toroidal flux, $s$ for quasisymmetric configurations (left) and non-quasisymmetric configurations (right)}
    \label{fig:gcvss}
\end{figure}

Nine configurations are represented in figure \ref{fig:gcvss} in the two plots showing $\Gamma_c$ as a function of normalized toroidal flux $s$. The six quasisymmetric configurations are plotted on the left, and the three non-quasisymmetric configurations are plotted on the right. Looking at the non-quasisymmetric configurations first, there is a clear distinction between the optimized configuration W7-X and the two LHD configurations. There is also a clear improvement between the LHD inward shifted configuration compared to the outward shifted configuration. However, the LHD inward shifted configuration has roughly the same magnitude of $\Gamma_c$ as the worst of the quasisymmetric configurations NCSX (note the difference in y-axis scale).

The quasisymmetric configurations also show considerable spread in the $\Gamma_c$ metric. Once again there is clear separation among the three quasiaxisymmetric configurations. The best performing case is the Simsopt equilibrium. Contrary to the quasisymmetry result, ARIES-CS outperforms NCSX with regard to the $\Gamma_c$ metric. This behavior is not surprising since the ARIES-CS optimization explicitly degraded quasisymmetry in order to improve energetic particle confinement \cite{mynick2006improving}. The particle loss results shown in figures \ref{fig:eloss} and \ref{fig:ploss} indicate that this optimization was successful. 

For the three quasihelically symmetric configurations the Wistell-A and Wistell-B configurations have almost identical values of $\Gamma_c$ (these are similar in scale to the W7-X value). The Ku5 configuration has a larger value in the core, but a lower value in the outer half of the plasma.  Since the Ku5 and Simsopt configurations represent the best performing configurations, it appears that the edge values in the outer half may be more important.  The importance of the quasisymmetric values on the outer half of the plasma has already been discussed with respect to optimizations of quasisymmetry \cite{henneberg2019improving} and the results presented here indicate that the values of $\Gamma_c$ in the outer half of the plasma may be more closely related to energetic particle confinement as well.

\begin{figure}
    \centering
    \begin{subfigure}{0.5\textwidth}
    \centering
    \includegraphics[width=\linewidth]{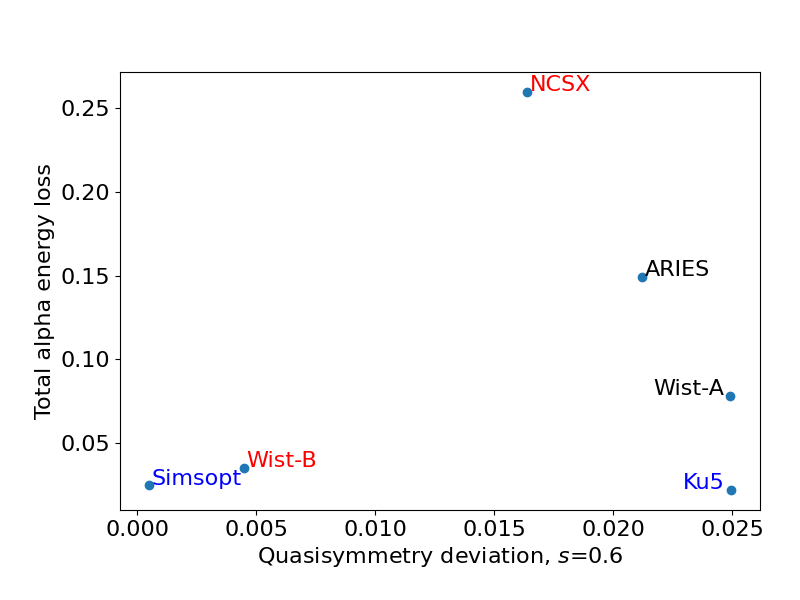}    
    \end{subfigure}%
    \begin{subfigure}{0.5\textwidth}
    \centering
    \includegraphics[width=\linewidth]{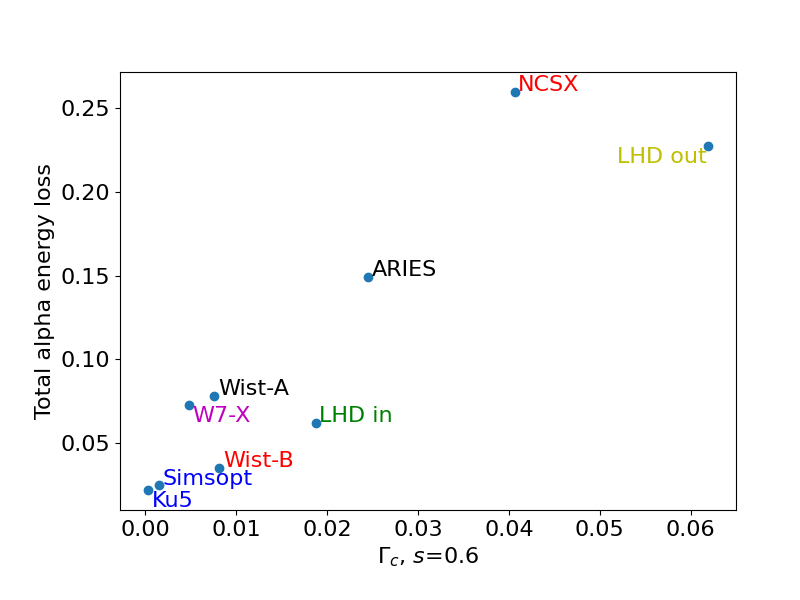}    
    \end{subfigure}%
    \caption{Values of total collisional energy lost as a function of quasisymmetry on the $s$=0.6 surface (left) and $\Gamma_c$ on the $s$=0.6 surface (right)}
    \label{fig:scatter}
\end{figure}

Figure \ref{fig:scatter} shows the total energy loss for each configuration plotted against the value of a parameter of interest evaluated at $s$=0.6. The deviation from quasisymmetry (for the QS configurations) is plotted on the left hand plot and $\Gamma_c$ is in the right hand plot. A correlation between alpha energy confinement and $\Gamma_c$ is clear from the right hand plot. Although a perfect correlation does not exist, it is clear that the best/worst performing configurations also have the best/worst performance with regard to this metric. For quasisymmetry the correlation is weaker. While the Simsopt and Wistell-B configuration both perform very well on this metric, the Ku5 configuration performs worse despite having the best overall energetic particle confinement.

Among the QA configurations the best performing configuration is the Simsopt QA, which also performs the best on both metrics, despite only optimizing for quasisymmetry.  As noted above, ARIES-CS performs worse in quasisymmetry but better in both $\Gamma_c$ and energetic particle confinement. 

The QH configurations include two stellarator configurations with excellent confinement, Ku5 and Wistell-B. For the QH configurations, Wistell-A performs worse in the quasisymmetry metric but approximately as well in $\Gamma_c$.  Both Wistell-A and Wistell-B were optimized including both quasisymmetry and $\Gamma_c$ in the target function.  The Ku5 configuration only optimized for quasisymmetry, but despite this has the lowest $\Gamma_c$. There is a caveat to the performances of the two high performing QH configurations.  In both the Wistell-B and Ku4 configurations, strong indentations in the plasma boundary make designing coils extremely challenging (see for example figure 13 in \cite{ku2010new}).  However, coils that reproduce the energetic particle properties have already been designed for the Wistell-A configuration \cite{bader2020advancing}.

The W7-X configuration was specifically designed for good energetic particle transport \cite{grieger1991physics} and indeed it outperforms ARIES-CS and is on par with the Wistell-A configuration.  Interestingly, the W7-X and Wistell-A configuration also have almost the same minimum value for $\Gamma_c$ so the points are very close in Figure \ref{fig:scatter}b.

\begin{figure}
    \centering
    \includegraphics[width=0.5\linewidth]{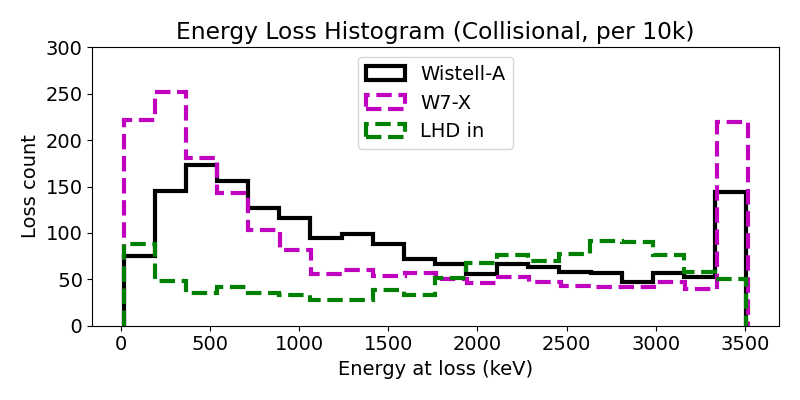}
    \caption{Histogram of number of particles lost (per 10k) as a function of the loss energy for Wistell-A (black), W7-X (magenta dashed) and LHD-inward shifted (green dashed).  Prompt losses at 3.5 MeV are at the far right side of the graph.}
    \label{fig:elosshist}
\end{figure}

The inward shifted LHD-like configuration has properties that deserve some attention.  The overall losses for this configuration are on par with both W7-X and Wistell-A.  Even more interesting is that this configuration does exceedingly well at confining prompt losses.  This good performance in the collisional results appears despite not performing particularly well on the $\Gamma_c$ metric. Figure \ref{fig:elosshist} illustrates this by plotting a histogram of the number of lost particles in the collisional calculation against the energy at which they are lost. Prompt losses are on the far right of the graph. These prompt losses are lowest for the LHD inward shifted configuration compared to both Wistell-A and W7-X. The LHD losses reach their maximum between 2.5 and 3.0 MeV after which they fall to a very low level. The Wistell-A and W7-X, in contrast have fewer particles lost between 2.5 and 3.0 MeV but considerably more particles lost at energies under 1 MeV. These configurations will be examined closer in sections \ref{ss:lhd} and \ref{ss:qh}.

\section{Discussion}
\label{sec:disc}

\subsection{Optimization}
One salient feature of the configuration scan is that two of the configurations with lowest achieved values of $\Gamma_c$ were not actually optimized for $\Gamma_c$ but rather for quasisymmetry only. These are the Ku5 and Simsopt configurations. Since perfect quasisymmetry will confine all particles, and a perfectly quasisymmetric configuration will have $\Gamma_c$ = 0, an optimization for quasisymmetry will almost always improve $\Gamma_c$ as well.  The cases where this does not occur, such as one of the configurations presented in \cite{bader2019stellarator} are fairly uncommon. In fact, in several steps of the Wistell-B optimization it was found that the best improvement on both metrics, quasisymmetry and $\Gamma_c$ was obtained when the $\Gamma_c$ optimization was turned off in the optimizer. 

The reverse is not true. Optimization for $\Gamma_c$ alone will almost never improve quasisymmetry. There are many pathways to improving $\Gamma_c$ that do not include quasisymmetry, for example, the optimizations that lead to W7-X.

Another important point of consideration is that many of the configurations were designed with additional metrics included. Both the NCSX and ARIES-CS equilibria placed strong emphasis on stability properties at finite $\beta$. Among other things, this generates a strong crescent shape at the $\phi = 0$ plane. In contrast the Simsopt QA is a vacuum configuration without a vacuum magnetic well and no attempt to provide stability at high pressure. The Simsopt configuration should be viewed as what is possible if you attempt to make the most quasiaxisymmetric configuration possible perhaps in opposition to other desired or even necessary properties. As always, significant effort is needed to weigh different optimizations considerations together to produce the ideal configuration for an experiment or reactor.

A similar story exists in the QH configurations. While Wistell-A and Wistell-B are both vacuum configurations, Wistell-A has a vacuum magnetic well and Wistell-B does not.  Furthermore, attempts to generate coils to reproduce the Wistell-A configuration were successful, while attempts to produce coils for Wistell-B or Ku5 have not been successful to date. Of course this does not mean that it is impossible to find coils for these configurations, just that it is comparatively easier to design coils for Wistell-A.  Despite having entirely different optimization schemes, Wistell-B and Ku5 have similar features. Specifically, both have a strong indentation in the teardrop shape that is very difficult to reproduce with coils. Future work that incorporates coil buildability should examine whether this feature is necessary for good confinement or not.

Finally we note that several of the configurations, namely the LHD configurations are actually built machines.  It is much easier to design a configuration with good parameters than to actually build one. It is possible that the performance of the other configurations would degrade due to accumulated errors in the construction process. Efforts to optimize taking into account manufacturing errors are being undertaken by others \cite{lobsien2020physics} and will not be discussed further here.

\subsection{LHD Inward Shifted}
\label{ss:lhd}

A surprising result from the configuration scan was the performance achieved by the LHD configuration. While it has been theorized and experimentally verified that the confinement improves in LHD with inward-shifted configurations, the actual performance deserves some additional discussion here.

\begin{figure}
    \centering
    \includegraphics[width=0.5\linewidth]{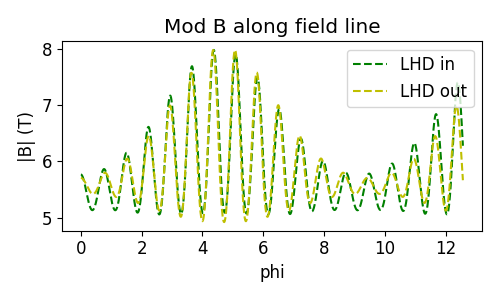}
    \caption{Magnitude of the magnetic field, $|B|$ along a field line for the LHD inward shifted (green) and outward shifted (yellow) configurations}
    \label{fig:lhdmodb}
\end{figure}

The specific optimization in question aligns the minimal values of the magnetic field on the surface, referred to as $\sigma = 1$ optimization. This can be seen in figure \ref{fig:lhdmodb} where a field line from both the outward and inward shifted cases are plotted as a function of toroidal angle. The minima align for the inward shifted case (green) but do not align for the outward shifted case (yellow). Another feature of these LHD-like equilibria is because the field is generated with helical coils, the field strength is smoothly varying with no local minima above the global minimum value. These local minima are problematic for particle confinement and can lead to promptly lost particles, similar to ripple trapped particles in a tokamak.

Since the maxima along the field line do not align, particles in the the LHD inward shifted configuration have a finite radial drift. In fact, as visible in figure \ref{fig:ploss} even in the absence in collisions, all trapped particles in both LHD configurations are eventually stochastically lost. The end result is a configuration which has no prompt losses due to the $\sigma=1$ optimization, but eventually loses all the particles. The parameters used for this calculation use the ARIES-CS parameters which have high plasma density and low temperature giving a slowing down time of $\approx$50 ms in the core, and considerably lower in the edge. When examining the collisional results under these conditions, the LHD inward shifted configuration compares favorably to configurations such as Wistell-A even though stochastic losses are very low in Wistell-A.

\subsection{QH: Collisionless vs Collisional losses}
\label{ss:qh}

The performance of Wistell-A is worth looking at closer.  In the collisionless losses (figure \ref{fig:ploss}) the total losses are low, below all other configurations except for Wistell-B and ITER. Furthermore, almost all the losses that do exist are prompt, occurring well before 1 ms. Yet, when collisions are added, Wistell-A performs significantly poorer to Wistell-B and Ku5 and instead performs equally well to W7-X which has more prompt losses and significantly more collisionless stochastic losses. The performance of Wistell-A is actually slightly worse than LHD, which has fewer prompt losses, but very large values of collisionless stochastic losses. 

\begin{figure}
    \centering
    \includegraphics[width=0.5\linewidth]{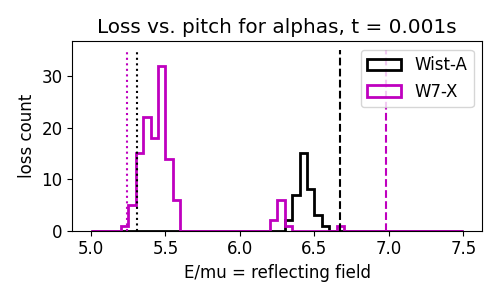}
    \caption{Histogram of collisionless prompt losses (less than 0.1 s) as a function of starting pitch for Wistell-A (black) and W7-X (magenta). The vertical dotted lines represent the minimum possible value of $E/\mu$ and the vertical dashed lines represent the trapped-passing boundaries for each configuration. The large population for W7-X corresponds to deeply trapped particles, while the other peak, larger on Wistell-A is near the trapped-passing boundary.}
    \label{fig:lvp}
\end{figure}

To understand this behavior it is necessary to not only distinguish between prompt and stochastic losses, but between the pitch angle of promptly lost particles. Particles can diffuse through phase space by pitch-angle scattering. Although pitch-angle scattering is small for 3.5 MeV alpha particles compared to momentum loss (by roughly a factor of 20), it still exists. If a particle diffuses into a region of phase space which is promptly lost, it will likely be lost before it can diffuse out. The distribution of these loss regions in phase space is important. If there is one major region of losses, such as all deeply trapped particles, the only particles that will be lost are those born in the region or close to it. However, if the prompt-loss regions are scattered around phase space, even if the total volume is lower, the amount of particles that may drift through a prompt-loss may be higher. Although verification will require statistical analysis tools beyond the scope of this paper, some basic analysis can be done by examining the pitches of promptly lost particles. Figure \ref{fig:lvp} shows a histogram of prompt (within 1 ms) collisionless particle losses for W7-X and Wistell-A as a function of pitch. The pitch parameter is given as $E/\mu$ where E is the particle energy and $\mu$ is the first adiabatic invariant. This ratio is the maximum field a particle can reach before reflecting. The trapped-passing boundary is slightly different for the configurations and is shown with vertical dashed lines. All the lost particles are trapped. Most of the losses from W7-X are from deeply trapped particles. All the losses for Wistell-A are near the trapped-passing boundary. While there are some losses near the trapped passing boundary for W7-X, these losses are significantly less in number than for Wistell-A.

One explanation for the relatively poor performance of Wistell-A compared to the expectations from collisionless losses is as follows. The LHD-inward shifted has no prompt losses and the collisional results appear similar to the collisionless results. Most particles are lost, but they are lost slowly. W7-X does have prompt losses, but these occur mostly in the deeply trapped particles. Only particles that are close to the deeply trapped region can diffuse onto lost orbits and be lost. In contrast, the losses from Wistell-A occur mostly near the trapped-passing boundary. The phase space volume near the trapped-passing boundary is significantly larger than the deeply trapped volume. For this reason, it is easier to diffuse into loss regions near the trapped-passing boundary and the losses are enhanced for Wistell-A when collisions are included. One result from this analysis is that if a configuration is to have prompt losses, it is far better to have them in deeply trapped regions.

\subsection{Limitations and Caveats}
The analysis presented in this paper is useful particularly for comparing different configurations, but to actually calculate losses in reactor configurations additional steps need to be taken. This section outlines some of the limitations of the calculations.

As noted above, the same profiles were used for every configuration despite significant differences in $\beta$. Furthermore, including a realistic profile for each configuration would obscure some of the differences between the configurations, which is the primary purpose of the results presented here.

Another limitation is the particle following algorithm is a guiding center algorithm and does not include finite gyro-orbits.  Finite orbits for alpha particles can be large and a full-orbit analysis between configurations would help determine whether the alpha loss estimates are accurate. At increased machine size, the effects of finite orbits are smaller and the guiding center approximation gets increasingly better. Many other effects are also not included, including any transport from Alfven Eigenmodes.

The configurations presented here all rely on \texttt{VMEC} equilibria. \texttt{VMEC} describes the equilibria as having nested toroidal flux surfaces without magnetic islands or stochastic field regions. This limitation is mitigated somewhat because the large orbits of alpha particles may average over small regions of stochasticity. Calculations with more realistic field, which also includes the effects from fields generated from coils are left for future work. 

Particles are considered lost if they pass beyond the penultimate surface in the \texttt{VMEC} equilibrium. In reality, particles may leave the confined plasma and reenter. This effect may be strongest in QA configurations which have the longest connection lengths and thus the largets banana widths.

\section{Conclusions and Outlook}
\label{sec:conc}
The analysis presented here shows that it is possible to optimize for stellarators to have good energetic particle confinement, often by ensuring very high quasisymmetry, as was done in the Ku5 and Simsopt equilibria. Post-hoc analysis of these two configurations indicate that it may be possible to achieve acceptable levels of energetic particle loss by optimizing for $\Gamma_c$ instead. Since $\Gamma_c$ is less restrictive than quasisymmetry, a large configuration space is available, and it may be possible to find a configuration that satisfies various other needs as well. Indeed, the Wistell-A configuration was optimized with $\Gamma_c$ and has both a vacuum magnetic well and a buildable coil set.

Unfortunately, none of the configurations presented here satisfy all of our needs for a stellarator reactor, which requires not only energetic particle confinement, but performance at high pressure, a buildable coil set, as well as other properties that are more difficult to quantify, like a viable divertor solution and reduced turbulent transport. As optimization algorithms and the physics metrics that feed into them improve, it is more likely that configurations which perform satisfactorily on all required axes will be found.

\section*{Acknowledgments}
The authors would like to acknowledge Mike Zarnstorff and Sam Lazerson for providing the NCSX and ARIES-CS equilibria, Don Spong for providing the ITER equilibrium, and Joachim Geiger and Carolin N\"uhrenberg for providing the W7-X equilibrium. Work for this paper was supported by DE-FG02-93ER54222, DE-FG02-00ER54546 and UW 2020 135AAD3116. Matt Landreman was supported by Simons Foundation (560651, ML)

\section*{References}
\bibliographystyle{iopart-num}
\bibliography{bader_iaea_2021}

\end{document}